\documentclass[journal]{IEEEtran}
\usepackage{graphicx}
\usepackage{epsfig}
\usepackage{dcolumn}
\usepackage{bm}
\usepackage{amsmath}
\usepackage{amssymb}
\usepackage{fancyhdr}
\usepackage{url}

\begin{document}

\title{Teaching Memory Circuit Elements via Experiment-Based Learning}
%
%
%

\author{Yuriy~V.~Pershin and Massimiliano~Di~Ventra
\thanks{Y. V. Pershin is with the Department of Physics
and Astronomy and USC Nanocenter, University of South Carolina,
Columbia, SC, 29208 \newline e-mail: pershin@physics.sc.edu.}
\thanks{M. Di Ventra is with the Department
of Physics, University of California, San Diego, La Jolla,
California 92093-0319 \newline e-mail: diventra@physics.ucsd.edu.}

}

%
%


\maketitle

\begin{abstract}
\boldmath
The class of memory circuit elements which comprises memristive, memcapacitive, and meminductive systems, is gaining
considerable attention in a broad range of disciplines. This is due to the enormous
flexibility these elements provide in solving diverse problems in analog/neuromorphic and digital/quantum computation;
the possibility to use them in an integrated computing-memory paradigm, massively-parallel solution of different optimization problems, learning, neural networks, etc. The time is therefore ripe to introduce these elements
to the next generation of physicists and engineers with appropriate teaching tools that can be easily implemented
in undergraduate teaching laboratories. In this paper, we suggest the use of easy-to-build emulators to provide a hands-on experience for the students to learn the fundamental properties and realize several applications of these memelements. We provide explicit examples of problems that could be tackled with these emulators that range in difficulty from the demonstration of the basic properties of memristive, memcapacitive, and meminductive systems to logic/computation and cross-bar memory. The emulators can be built from off-the-shelf components, with a total cost of a few tens of dollars, thus providing a relatively inexpensive platform for the implementation of these exercises in the classroom. We anticipate that this experiment-based learning can be easily adopted and expanded by the instructors with many more case studies.
\end{abstract}


%
\IEEEpeerreviewmaketitle

\section{Introduction}

\IEEEPARstart{M}emory circuit elements (memelements) \cite{diventra09a} are resistors,~\cite{chua71a,chua76a} capacitors and inductors~\cite{diventra09a} with memory. These are two-terminal {\it dynamical} non-linear circuit elements that typically retain information even without a power source~\cite{pershin11a}. As such, they are of great importance in solid-state memory technologies~\cite{Green07a,Karg08a,Sawa08a} since they may replace
conventional Flash memory~\cite{pershin11a}. Their potential, however, is not limited to
such application. Indeed, their intrinsic non-linear dynamics, coupled with information-processing capabilities,
makes them ideal candidates for a wide range of tasks, ranging from massively-parallel
solution of optimization problems~\cite{pershin11d} to neuromorphic circuits~\cite{pershin10c,jo10a,Choi09a,Lai10a,Alibart10a,fontana10a}, digital computation on the {\it same} platform as memory storage~\cite{strukov05a,Snider07a,borghetti10a,pershin10e}, enhancement of quantum computing processing~\cite{pershin10e}, and even understanding of biological processes~\cite{pershin09b,Johnsen11a}.

Since the use of these {\it memelements} in research is increasing rapidly and it has been announced that the memristive
random access memory may reach the market already as early as 2013 \cite{reram2013}, it is our opinion that their introduction into the electrical engineering and physics university curricula is a {\it sine qua non} for the training of the next generation of scientists and engineers worldwide. In this respect, their theoretical discussion as part of a regular circuit theory course would already be of great value. However, as it is often the case, a hands-on experience would provide
an added benefit to the students which can be cherished for years to come.

Here, a serious difficulty arises: memristive, memcapacitive, and meminductive systems are not yet commercially available. And even when some of these will become available, they may require handling with
care, and thus will have a limited time span in the student laboratory.

\begin{figure}[b]
 \begin{center}
 \includegraphics[angle=0,width=6.5cm]{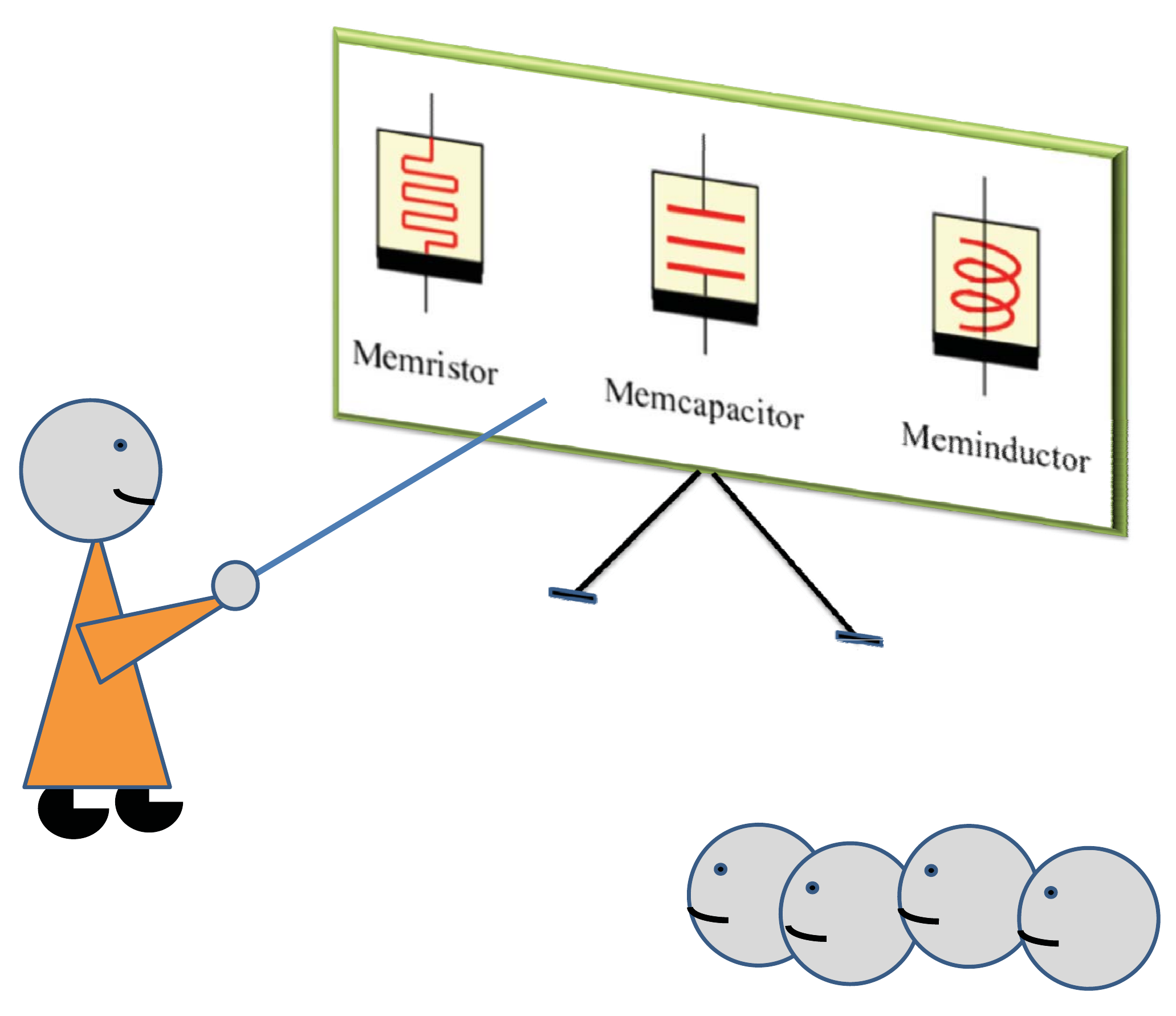}
\caption{\label{fig01} Teaching memory circuit elements. Symbols of memristive, memcapacitive and meminductive systems
are shown on the blackboard.
Modified with permission from \cite{pershin10e}. Copyright © (2011) by the IEEE.
}
 \end{center}
\end{figure}

Borrowing from our own personal experience, we then suggest to use memristor~\cite{pershin10d}, memcapacitor and meminductor \cite{pershin11b} {\it emulators}, which we have developed for research purposes, as the main tools to complement the theoretical teaching of memelements~\cite{Pazienza11a}, see Fig. \ref{fig01}. A few other known designs of emulators~\cite{chua71a,pershin09e,biolek10a,biolek11a,biolek11b,Wang2011a} can be alternatively employed, but the ones we report in this paper combine relative simplicity with versatility and close adherence to the fundamental properties of memelements.

We then suggest a sequence of laboratory projects based primarily on our own recent research work. This sequence
will guide the students from the basic properties of memelements to a wide range of applications with increasing
complexity. For each project, we also provide a list of equipment required for its execution. From this it will be clear that all the projects do not require more than basic off-the-shelf components and tools available in most
of electrical engineering and physics undergraduate laboratories. Obviously, there are many more examples that can be
implemented based on these emulators platforms. We leave this to the imagination and skills of both the lecturers and
the students.

In addition, to facilitate the preparation of the suggested laboratory
exercises by the instructors we have posted the documentation necessary
to build the memristor emulator - upon which all other emulators can be
derived - at the following web page \cite{webpage}. In there, we release the scheme, the
printed circuit board (PCB) layout of the memristor emulator and the
code (in C language) for the micro-controller.

The paper is organized as follows. In Sec.~\ref{def} we introduce the general definitions of memory circuit elements and outline their main properties. Electronic emulators implementing all memory circuit elements (memristive, memcapacitive, and meminductive systems) are presented in Sec.~\ref{emulate}. Sec. \ref{projects}
contains a sequence of projects  of increasing complexity and difficulty that includes experiments with basic properties of memelements and hysteresis loops, learning circuits, programmable analog circuits, memristive neural networks, logic circuits and cross-bar array memory. In Sec. \ref{concl} we conclude with
some final remarks and considerations on the proposed teaching approach.

\begin{figure}[tb]
 \begin{center}
 \includegraphics[angle=0,width=6.5cm]{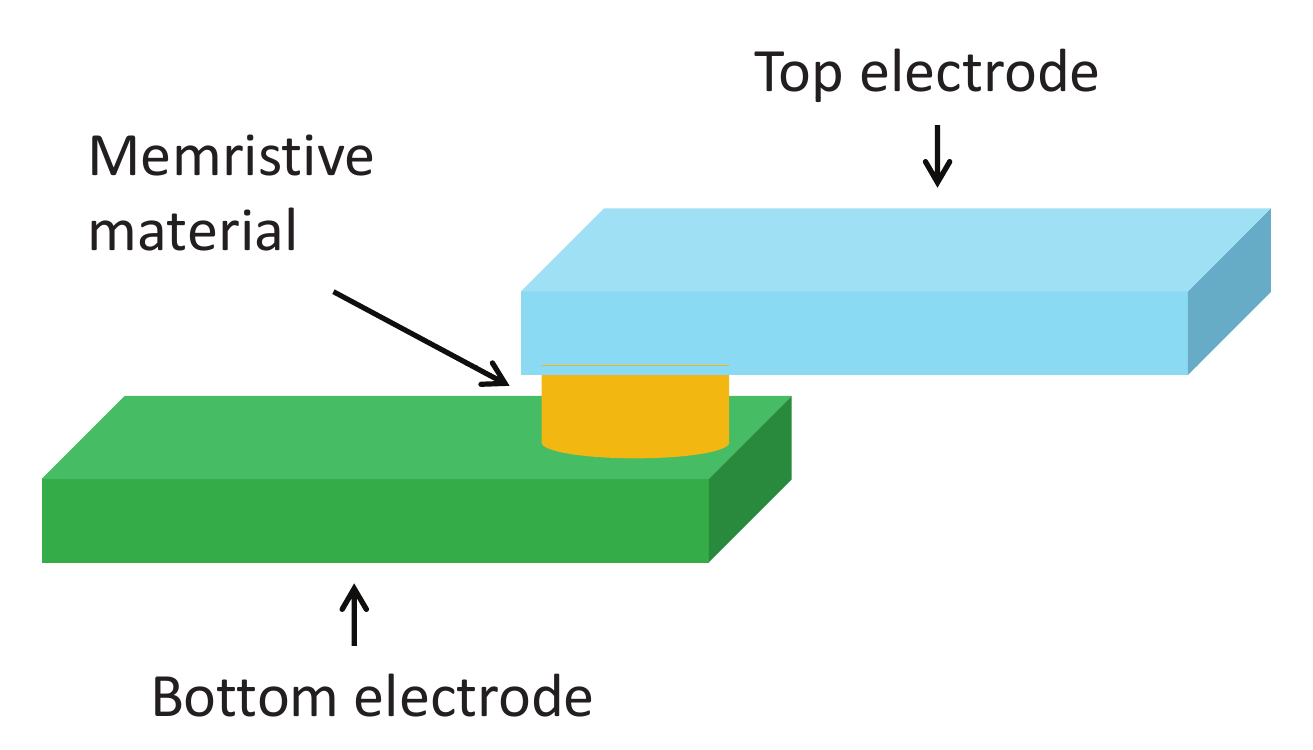}
\caption{\label{fig02} Schematic of a resistance-switching memory cell. The state of the device
is determined by the resistance of the memristive material sandwiched between two metal electrodes.
The metal electrodes are typically separated by a few tens of nanometers (see, e.g., Refs. \cite{yang08a,jo09a,waser07a}).}
 \end{center}
\end{figure}

\begin{figure}[tb]
 \begin{center}
 \includegraphics[angle=0,width=6.5cm]{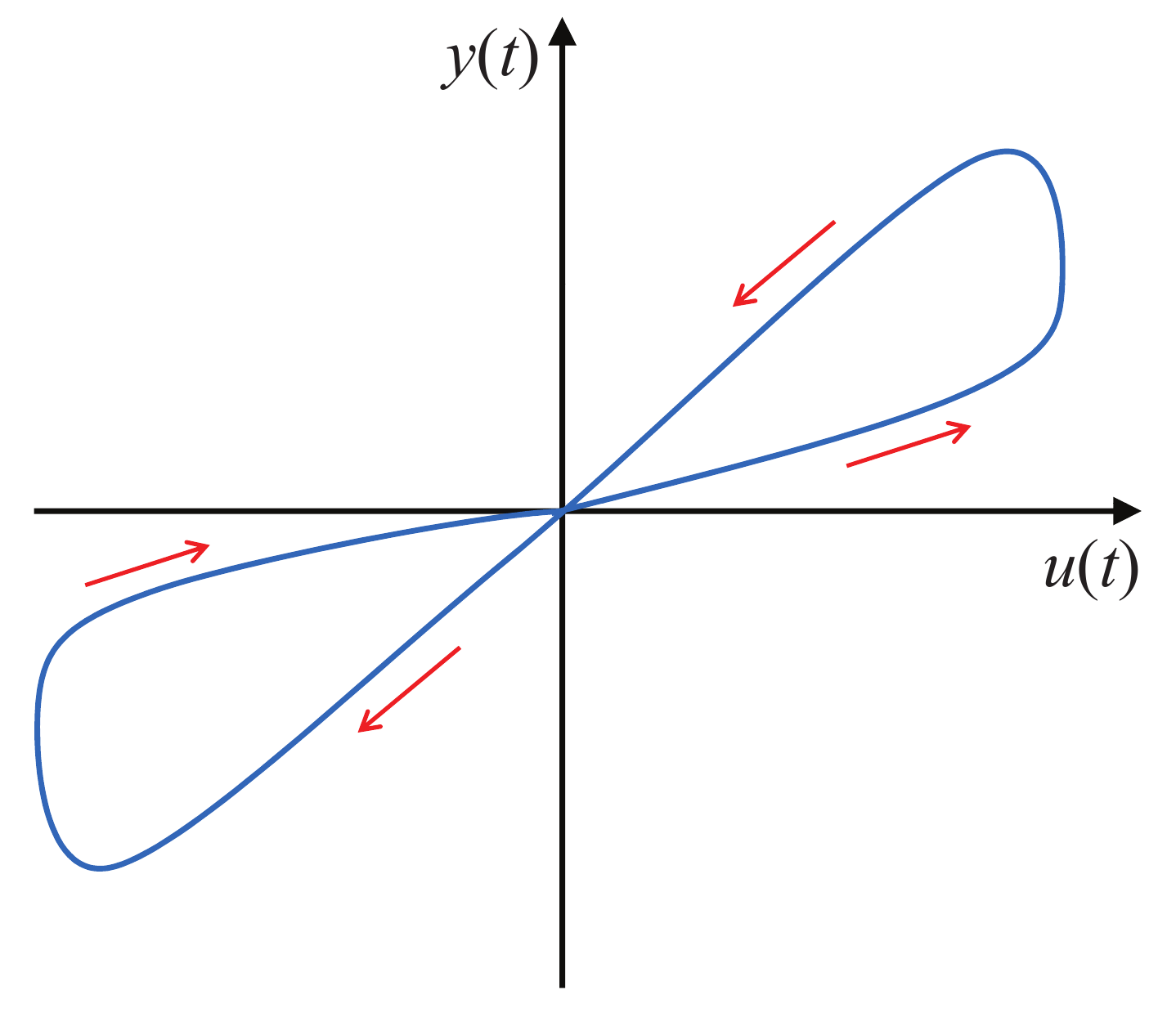}
\caption{\label{fig03} Pinched hysteresis loops are common distinctive feature of memory
circuit elements. Such loops are frequency-dependent and obtained when the input circuit variable $u(t)$ in Eq. (\ref{Geq1}) is periodically driven. The arrows show the direction along the loop. In this particular example of bipolar switching~\cite{pershin11a},
the response function $g$ increases at positive values of $u$ and decreases when $u$ is negative.}
 \end{center}
\end{figure}

\section{Definitions of Memory Circuit Elements}\label{def}

\begin{figure}[b]
 \begin{center}
 \includegraphics[angle=0,width=7.5cm]{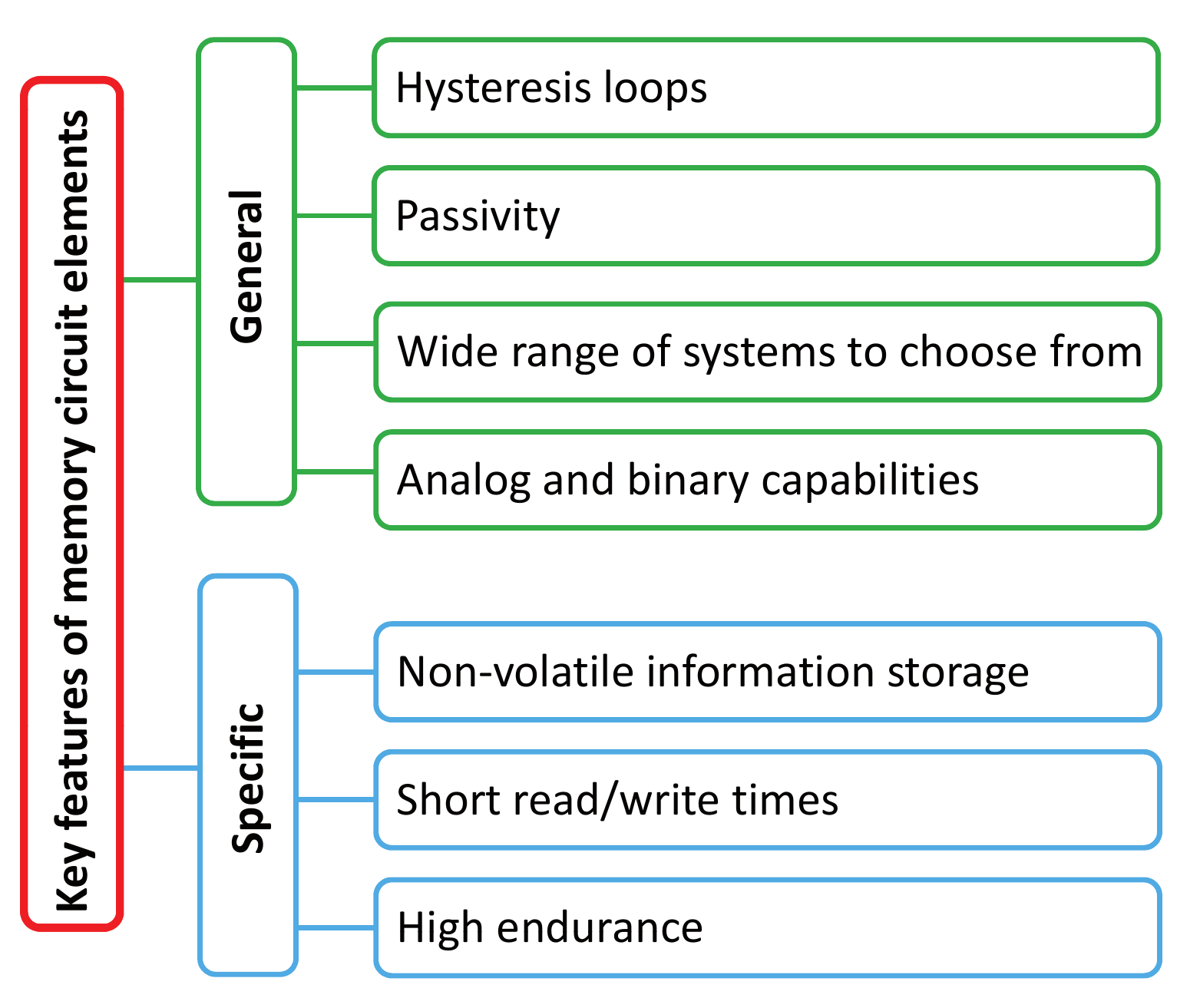}
\caption{\label{fig04} Some key properties of memory circuit elements separated into general (found almost in any memory circuit element) and specific (definitely not general, but present in some) categories.}
 \end{center}
\end{figure}

The general axiomatic definition of memelements is straightforward and it has been reported in the original
publications~\cite{chua71a,chua76a,diventra09a}. Here, we follow Ref.~\cite{diventra09a} considering constitutive
relations formed by pairs of fundamental circuit variables $u(t)$ and $y(t)$ (i.e., current,
charge, voltage, or flux). In these relations, the pairs of circuit variables are coupled by
the response functions, $g$, depending also on
a set of $n$ state variables, $x=\{x_i\}$, describing
the internal state of the system. These variables could be related, e.g., to the electric \cite{martinez09a} or spin \cite{pershin08a,wang09a} polarization, system geometry \cite{pershin11c}, phase state \cite{Wright11a}, temperature \cite{chua76a} or some other properties. The resulting memory element is then described by
the following relations~\cite{diventra09a}
\begin{eqnarray}
y(t)&=&g\left(x,u,t \right)u(t) \label{Geq1}\\ \dot{x}&=&f\left(
x,u,t\right) \label{Geq2}
\end{eqnarray}
where $f$ is a continuous $n$-dimensional vector function, and we
assume on physical grounds that, given an initial state $u(t=t_0)$
at time $t_0$, Eq.~(\ref{Geq2}) admits a unique
solution. If $u$ is the current and $y$ is the voltage then Eqs. (\ref{Geq1}), (\ref{Geq2}) define memory resistive (memristive) systems. In this case $g$ is the {\em memristance} (for memory resistance). In memory capacitive (memcapacitive) systems, the
charge is related to the voltage so that $g$ is the {\em memcapacitance} (memory capacitance); while in memory inductive (meminductive) systems the flux is related to the current with $g$ the {\em meminductance} (memory inductance).
The remaining three pairs of fundamental circuit variables
do not give rise to new devices:
the pairs charge-current and voltage-flux are linked
through equations of electrodynamics, and devices defined by the relation of charge and
flux (which is the integral of the voltage) are not considered as
a separate group since such devices can be redefined in the
current-voltage basis~\cite{chua71a}. Strictly speaking, memristors, memcapacitors and meminductors are 
are special {\it ideal} instances of memristive, memcapacitive and meminductive systems~\cite{diventra09a}, respectively~\footnote{Many researchers use the terms memristor and memristive system, memcapacitor and memcapacitive system, and meminductor and meminductive system interchangeably.}.

\begin{figure*}[tb]
 \begin{center}
 \includegraphics[angle=0,width=14cm]{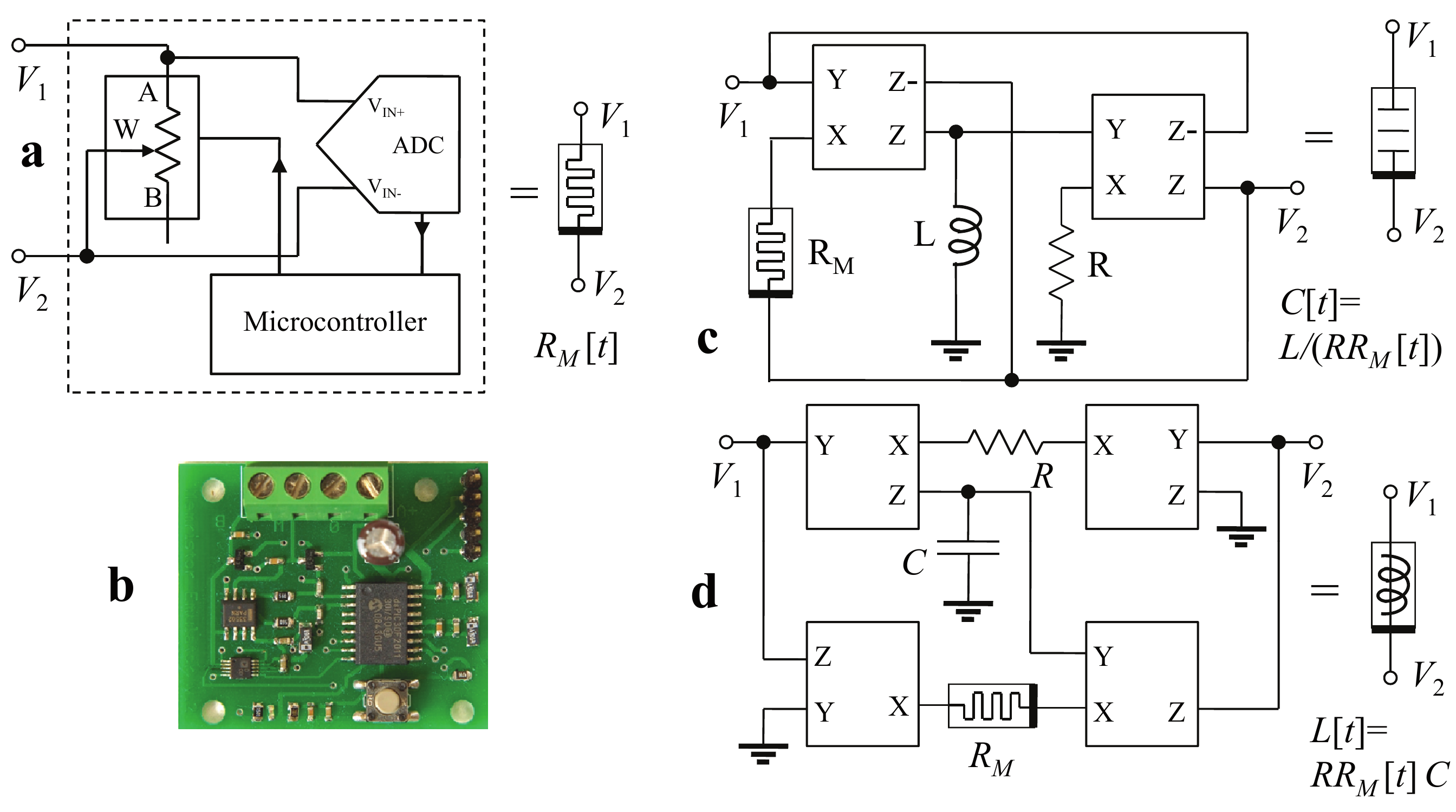}
\caption{\label{fig1} Circuits simulating a memristive system ("memristor emulator") {\bf a},  memcapacitive system {\bf c} and
meminductive system {\bf d}. Corresponding symbols of memory circuit elements are shown to the right of each circuit. {\bf b} Image of an experimental realization of memristor emulator. {\bf a} is reprinted with permission from \cite{pershin10d}.
Copyright © (2010) by the IEEE. {\bf c} and {\bf d} are reprinted with permission from \cite{pershin11b}. Copyright © (2011) by the IET.}
 \end{center}
\end{figure*}

As an example of a memory circuit element defined by Eqs. (\ref{Geq1}), (\ref{Geq2}) let us consider a current-controlled memristive system \cite{chua76a,diventra09a}. It is specified as
\begin{eqnarray}
V_M(t)&=&R\left(x,I,t \right)I(t), \label{eq1}\\
\dot{x}&=&f\left(x,I,t\right), \label{eq2}
\end{eqnarray}
where $V_M(t)$ and $I(t)$ denote the voltage and current across the
device, respectively, and $R$ is the memristance. It clearly follows from Eqs. (\ref{eq1}),
(\ref{eq2}) that the memory feature of memristive systems (as well as of all other memelements)
is provided by the internal state variable(s) $x$.
Definitions of all the other elements are easily derived by considering the different constitutive variables.

Fig. \ref{fig02} shows a typical structure of
a resistance-switching memory cell. It has been shown experimentally that many different material combinations in the geometry of Fig. \ref{fig02} exhibit the resistance switching effect, such as, for example, Pt-TiO$_2$-Pt~\cite{yang08a} and (p-Si)-(a-Si)-Ag~\cite{jo09a} structures. For more detailed information about experimental systems showing memristive, memcapacitive and meminductive properties, we refer the reader to our recent review paper~\cite{pershin11a}.

Memory circuit elements are characterized by a typical "pinched hysteresis
loop" in their constitutive variables when subject to a periodic input (see Fig. \ref{fig03}), their characteristics (memristance, memcapacitance and meminductance) vary between two limiting values (with exceptions as discussed in Refs.~\cite{pershin11a,martinez09a,krems2010a}), and may depend on initial conditions~\cite{Corinto11a}. The hysteresis is generally more pronounced at frequencies of the external input that are comparable to frequencies of internal processes that lead to memory. In threshold-type memdevices, however, the hysteresis is also significant at low frequencies. In many cases, at very low frequencies memory elements behave as non-linear elements while at high frequencies as linear elements.

In addition, the hysteresis loops can be with and without self-crossing (type-I and type-II crossing behavior) \cite{pershin11a} and, in many systems, the internal state variable remains unchanged for a long time without any input signal applied, thus
providing non-volatile memory. Importantly, memristive systems are always dissipative elements, while equations describing memcapacitive, and meminductive systems may describe non-dissipative and dissipative (and in principle also active) behavior   \cite{diventra09a}. In Fig.~\ref{fig04} we also summarize some key properties of memory circuit elements which make them very appealing for applications. In particular, while non-volatile information storage is not a necessary condition for the definition of the various memelements, it is nonetheless a desirable property for several applications.

It is also worth mentioning that the state variables - whether from a continuum or a discrete set of states - may follow a {\it stochastic differential equation} rather than a deterministic one \cite{pershin11a}. This case has received much less attention even though
interesting phenomena, such as noise-induced hysteresis have been predicted \cite{Stotland11a}. Although we will not consider it in this paper, this case can be easily implemented in the emulators we discuss below,
providing additional teaching tools at the instructor's disposal.

\section{Emulators}\label{emulate}

Electronic schemes of emulators can be classified into analog (e.g., as in Ref. \cite{chua71a}) and digital \cite{pershin10d,biolek11b} ones. The digital approaches \cite{pershin10d,biolek11b} are more flexible since they are based on a microcontroller whose program parameters/model can be easily adjusted. Fig. \ref{fig1}{\bf a} presents the design of a microcontroller-based memristor emulator \cite{pershin10d}. Its main parts include a microcontroller, digital potentiometer and analog-to-digital converter. Via the analog-to-digital converter (ADC), the microcontroller cyclically reads the voltage applied to the digital potentiometer, calculates an updated value of memristance (using pre-programmed equations of voltage-controlled or current-controlled memristive system, see Eqs. (\ref{eq1}), (\ref{eq2})) and writes the updated value of memristance into the digital potentiometer.

We have recently re-designed our initial version of memristor emulator \cite{pershin10d} and have utilized the novel version in experiments on chaotic behavior of certain memristive systems \cite{driscoll11a}. The hardware documentation and internal microcontroller program written in C of the re-designed version can be found at \cite{webpage}. The total cost of this emulator is about 25 USD.

\begin{figure}[tb]
 \begin{center}
 \includegraphics[angle=0,width=7cm]{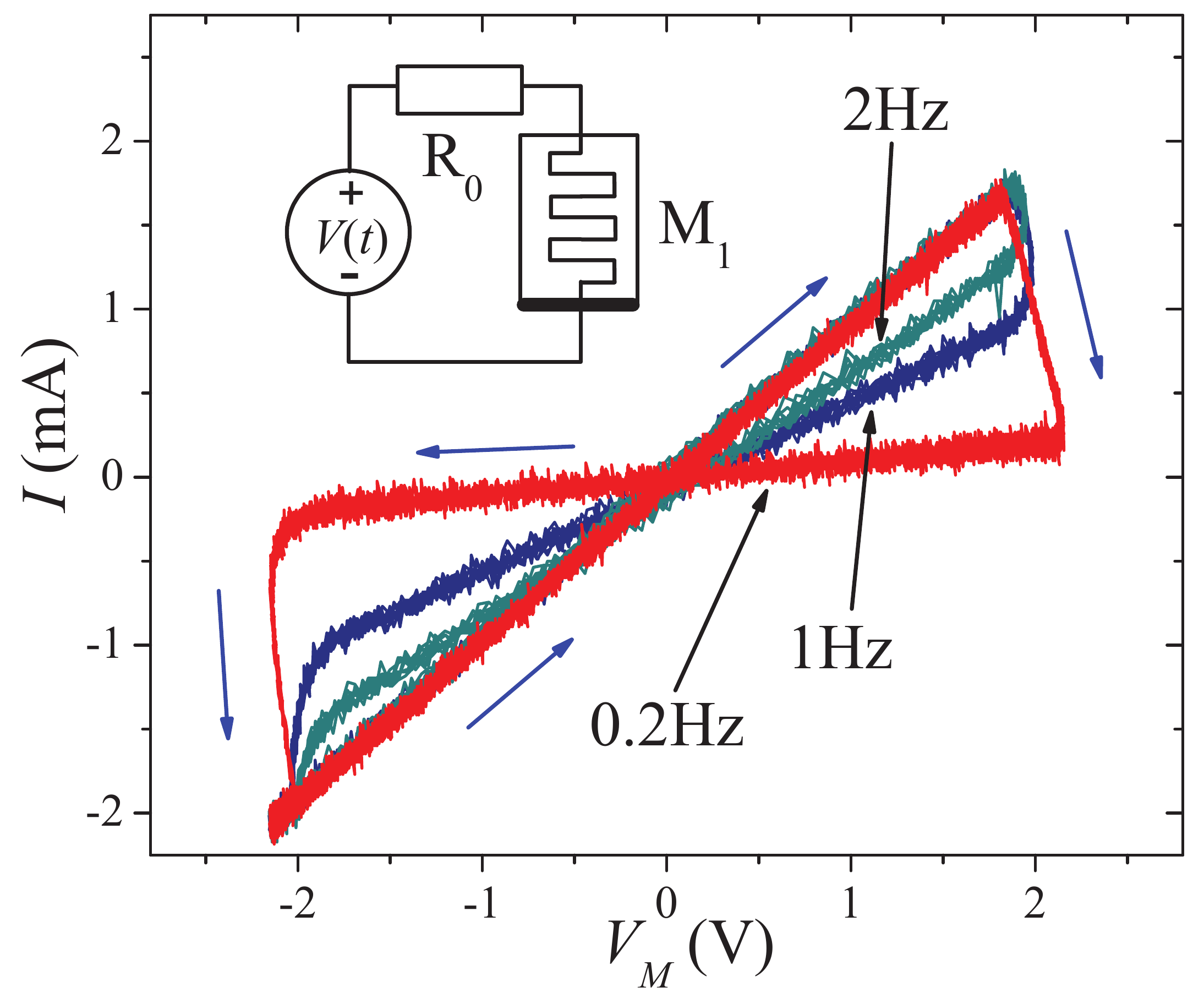}
\caption{\label{fig2} Pinched hysteresis loops obtained with a memristor emulator at different frequencies of the applied ac-voltage (see Ref. \cite{pershin10d} for more details).
The voltage drop over a small value resistor $R_0 =100\Omega$
in the inset was used to measure the current. The memristance of the emulator was limited between 1k$\Omega$ and 10k$\Omega$. Adapted with permission from \cite{pershin10d}.
Copyright © (2010) by the IEEE.}
 \end{center}
\end{figure}

Emulators of memcapacitive and meminductive systems can be developed based on the memristor emulator. Fig. \ref{fig1}{\bf c},{\bf d} show possible realizations of such emulators using current conveyors (see Ref. \cite{pershin11b} for more details). It is important that in these circuits the external terminals are floating, namely, any voltage can be applied to these terminals. Simpler emulators of memcapacitive and meminductive systems utilize a single operational amplifier \cite{pershin09e}. However, their application is limited to the cases when a memcapacitive or meminductive system is connected through a resistor to the ground (with a single floating terminal).
The total cost of these emulators is just slightly above the cost of the memristor emulator since memcapacitor and meminductor emulators include few additional
circuit components (see Fig.~\ref{fig1}).

\section{Teaching methodology} \label{projects}

\subsection{Prerequisites and pedagogical issues}
We expect the students to have already passed one course (e.g., an electromagnetism course in physics) where the basics of linear electronics are presented. Any knowledge of embedded electronics (needed to understand details of operation and of control program of the memristor emulator) may help but is not required. The knowledge of basic programming languages (e.g., C or Fortran) is also not necessary, although it would be of added
benefit to the students. Memristor, memcapacitor and meminductor emulators can then be introduced as "black boxes" with some initially pre-defined properties that, first of all, should be measured and understood and, after that, used in electronic circuits. Such an approach will be suitable for undegraduate students from different disciplines and is based on minimal prerequisite requirements.

On the other hand, students from EE departments - and we expect students of some Physics departments as well - that are already familiar with embedded electronics can use this advantage in order to understand how the memristor emulator works and write their own codes mimicking a variety of memristive systems. In this case, the very first task we would suggest is to present the students to the hardware of the memristor emulator (describing both the independent operation of its three main parts: the micro-controller, the analog-to-digital converter, and the digital potentiometer, as well as their combined operation).

If the instructors do not think the students can write their own code for the micro-controller, then they should provide the code to the students with some
basic understanding of its structure, function and, most importantly, direct them to the point where they can change at least the lines related to the memory model implemented - the function $f$ in Eq.~\ref{eq2}. This way, the micro-controller code (at least) partially looses its "black-box" status, thus allowing the students to experiment with different models of memory.

Most importantly, it will be easier for the students to understand the new concept of memelements based on already known concepts (elements of traditional electronics) if the letter ones are discussed as specific
limiting cases of memelements. Finally, it goes without saying that it is expected the instructors to collect feedback from the students to be aware and promptly respond to possible issues that may arise during both class lectures and
laboratory experiments.

\subsection{Basic properties of memory circuit elements and hysteresis loops}

{\it Class Lecture:} The instructor should devote an entire lecture to the general definition of memory elements, their main properties, and put them into a wider context by showing a wide
variety of experimental devices that can be described as - or as a combinations of - memristive, memcapacitive, and meminductive systems. This initial lecture should also stress the fundamental difference with standard circuit elements and how these
are recovered from the memelements definition. It would also be beneficial for the student a brief and general introduction to the possible applications of these elements, which anticipates the different
experimental projects in the series.

{\it Laboratory Experiment:} The first suggested project in the sequence is related to the basic understanding of the response of memory circuit elements to ac and dc applied signals.
Specifically, it is important to realize that parameters of many memristive, memcapacitive and memindcutive systems change between
two limiting values when the magnitude of the applied signal exceeds a threshold value \cite{pershin11a}.

In addition, as discussed previously, a distinctive feature of many memory circuit elements is a frequency-dependent pinched hysteresis loop \cite{chua76a,diventra09a,pershin11a}, see Fig. \ref{fig2}. Pinched hysteresis loops are
graphically obtained when the output circuit variable is plotted versus the input circuit variable for a periodically-driven device. An example of such pinched hysteresis loops is depicted in Fig. \ref{fig2} showing that the loop span depends on the applied voltage frequency. In this project,
students could be asked to measure limiting values of memristance, memcapacitance and meminductance, threshold voltages, and record hysteresis loops
at different frequencies of the applied signal. This project requires emulators of memory circuit elements, ac- and dc-voltage sources and data acquisition system.

\subsection{Learning circuit}

\begin{figure}[t]
 \begin{center}
 \includegraphics[angle=0,width=6cm]{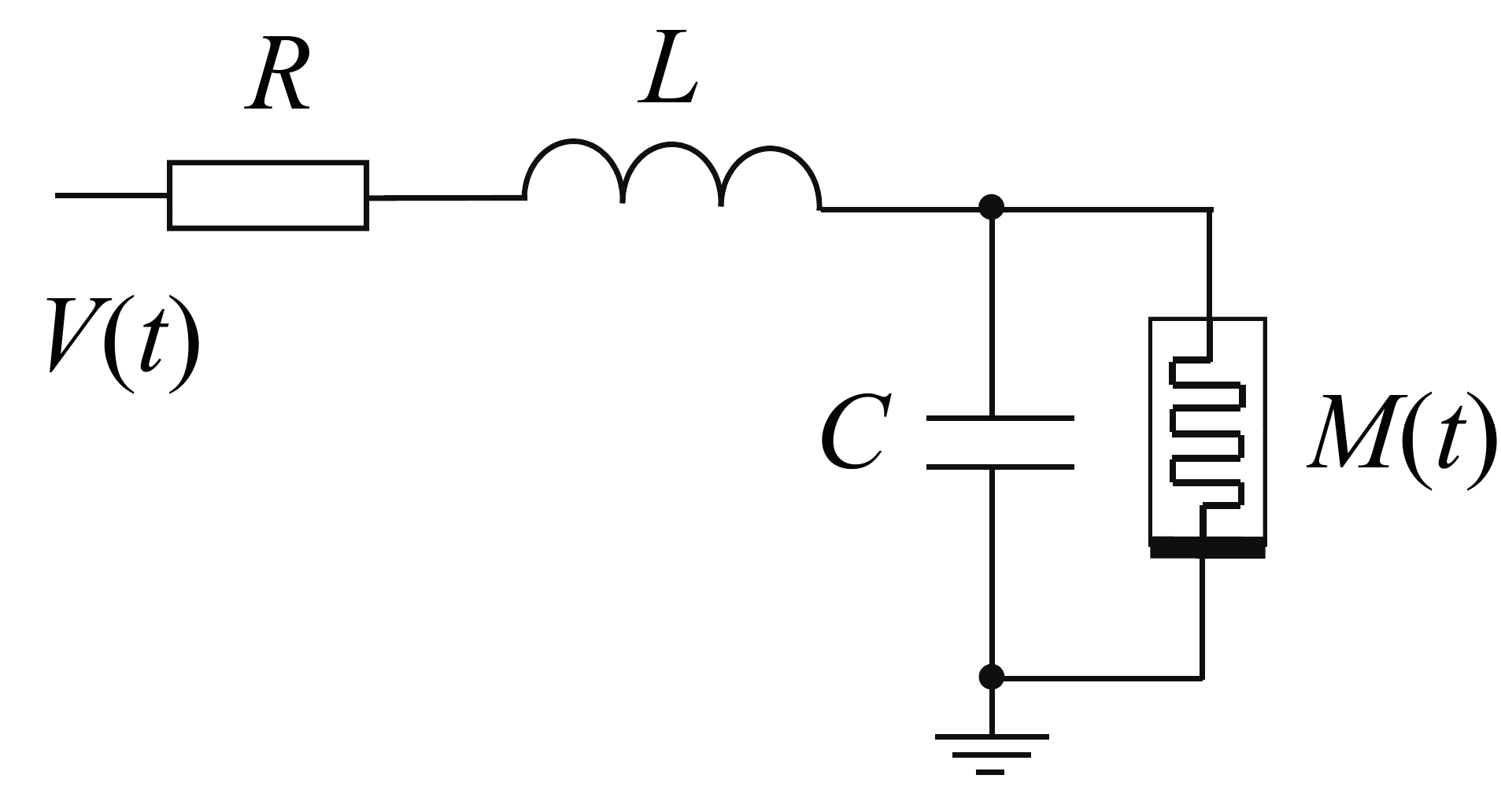}
\caption{\label{fig7} Learning circuit \cite{pershin09b} mimicking the adaptive behavior of {\it Physarum Polycephalum} \cite{saigusa08a}.}
 \end{center}
\end{figure}

{\it Class Lecture:} A general lecture on the "adaptive" behavior of these elements to input waveforms would introduce the students to the notion that memelements can adjust and retain {\it analog} information, namely, when
combined in circuits, they provide a flexibility not encountered in standard circuit elements. Here, the instructors may want to discuss that these elements are indeed ideal to model the adaptive behavior and learning of biological systems~\cite{pershin09b,pershin10c,pershin10e,Johnsen11a}. For instance, similar to changes in behavior of biological organisms due to varying environmental conditions, the state of a memory circuit
element changes when an external signal is applied. The students should then acquire an even deeper understanding of the information-storage capabilities of memelements with connections
to fields apparently unrelated to electronics, such as animal and human learning.

{\it Laboratory Experiment:} Using the above analogy, a "learning circuit" as that depicted in Fig. \ref{fig7} \cite{pershin09b} can be an ideal candidate to show analogies with the adaptive behavior of unicellular
organisms, such as amoebas \cite{saigusa08a}. The main idea of the circuit operation is related to a change of the
damping rate of an LC-contour when the ac-voltage frequency approaches the resonant frequency of the contour. Experimentally, these learning circuits - which operate also as adaptive filters - have been demonstrated using VO$_2$ as memory material \cite{driscoll11a}.

For the specific project, a learning circuit can be built using either a memristor emulator (as described in
Fig.~\ref{fig7}) or a memcapacitor one. Students could be asked to study the learning scheme from Fig. \ref{fig7} and design a different learning circuit based on a memcapacitor emulator. This project requires memristor and memcapacitor emulators, basic circuit elements, ac-voltage source and data acquisition system.

\subsection{Programmable analog circuits}

\begin{figure}[t]
 \begin{center}
 \includegraphics[angle=0,width=7cm]{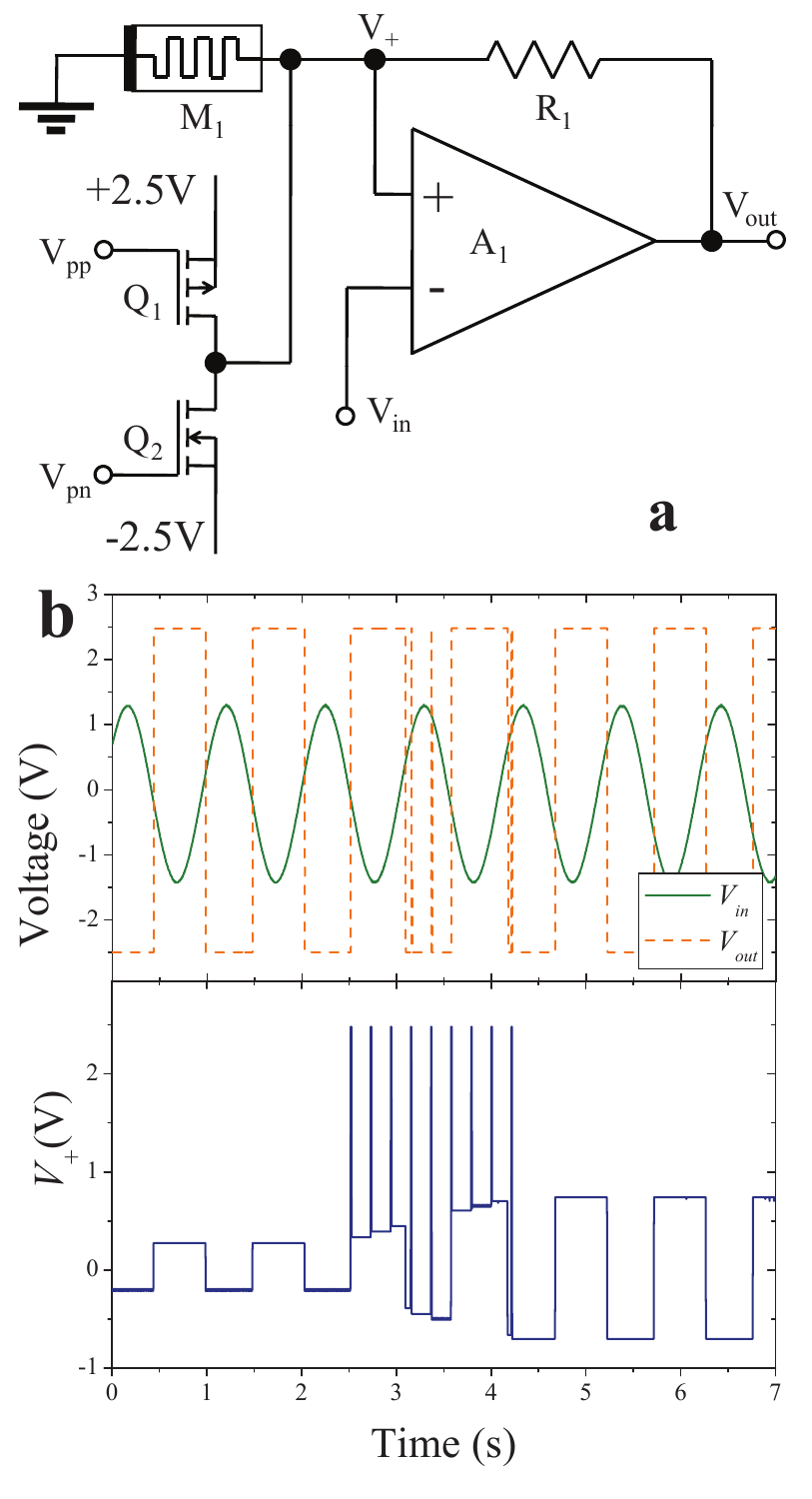}
\caption{\label{fig3} An example of programmable analog circuit with memory circuit elements:
{\bf a} memristor-based programmable switching thresholds Schmitt trigger. {\bf b} Programmable switching thresholds
Schmitt trigger response to the input voltage $V_{in}=V_0\sin
(2\pi f t)$ with $V_0=1.3$V and $f=1$Hz, and several positive
programming pulses of 10ms width applied in the time interval
between 2 and 4 seconds. As it is evident, the increase of memristance M$_1$ due to the programming pulses results in the trigger's
threshold increase. Reprinted with permission from \cite{pershin10d}.
Copyright © (2010) by the IEEE.}
 \end{center}
\end{figure}

\begin{figure}[t]
 \begin{center}
 \centerline{
    \mbox{{\bf a}}
    \mbox{\includegraphics[width=5.00cm]{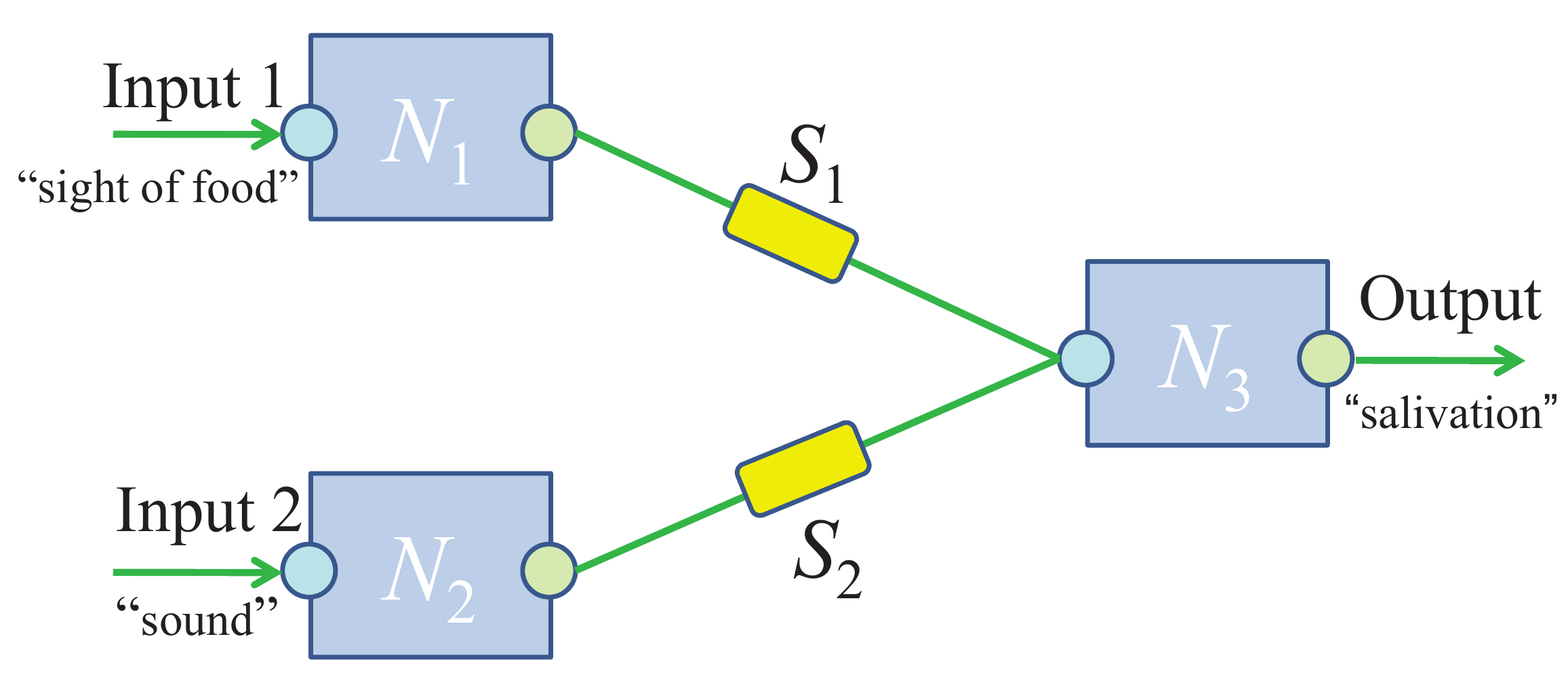}}
 }
 \centerline{
    \mbox{{\bf b}}
    \mbox{\includegraphics[width=7.00cm]{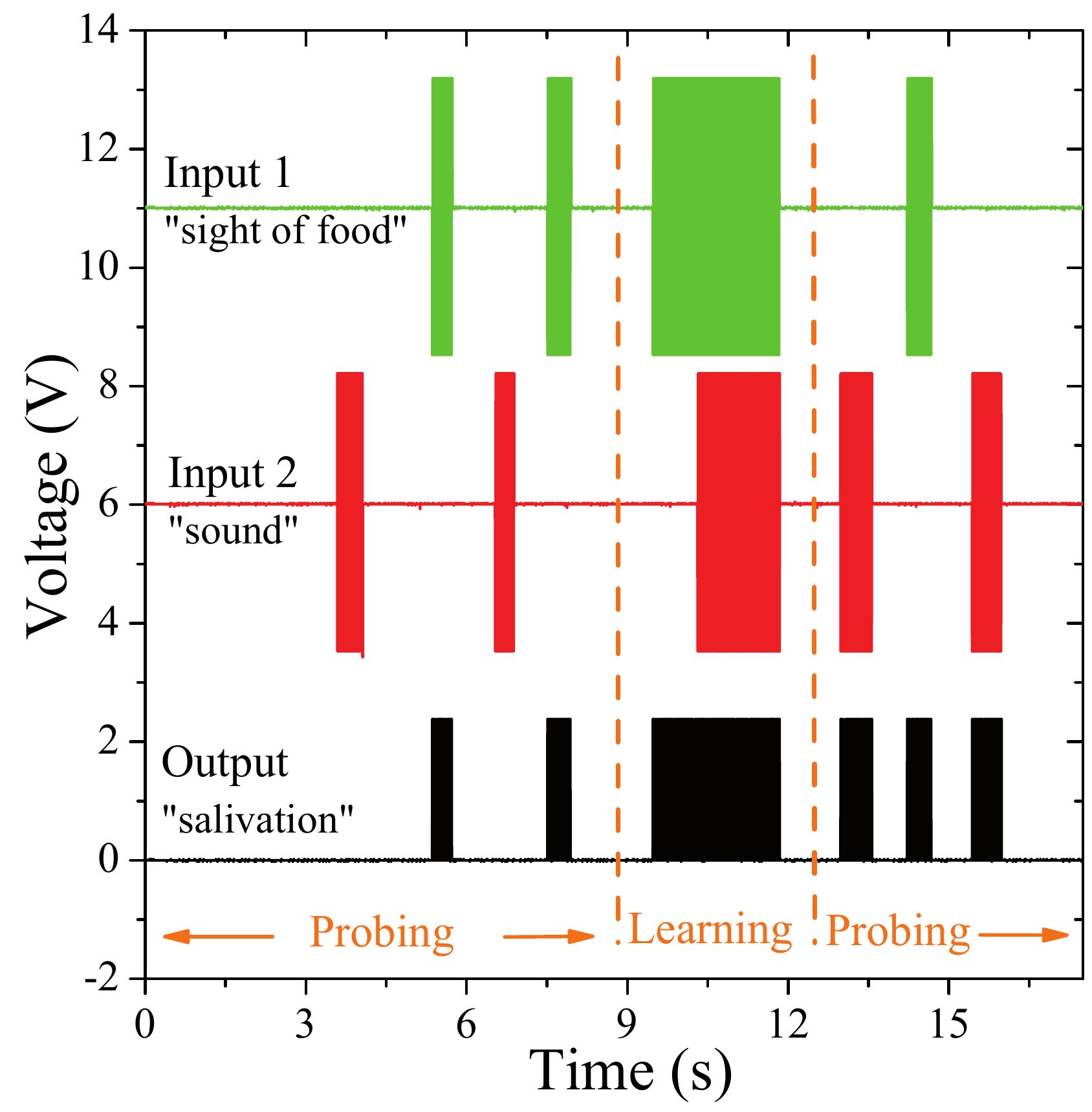}}
  }
\caption{\label{fig4} {\bf a} A neural network containing three electronic neurons (N$_1$, N$_2$ and N$_3$) connected by two memristive synapses (S$_1$ and S$_2$). {\bf b} Demonstration of associative memory in the neural network sketched in {\bf a} (for details, see Ref. \cite{pershin10c}).
Initially, only a signal from N$_1$ neuron activates the output neuron as it is shown in the first "probing" phase ($S_1$ is in ON state and $S_2$ is in OFF state). The association of the Input 2 signal with the Output develops in the "learning phase" when
N$_1$ and N$_2$ neurons are simultaneously activated. In this case, a signal at the Input 1
excites the third neuron that sends back-propagating pulses of a negative
amplitude. These pulses, when applied simultaneously with forward propagating pulses from the Input 2 to the second memristive synapse S$_2$ cause it to learn ($S_2$ memristive synapse switches into the ON state). The final "probing" phase demonstrates that signals from both N$_1$ and N$_2$ activate the output neuron.
Reprinted from \cite{pershin10c}. Copyright © (2010) with permission from Elsevier.}
 \end{center}
\end{figure}

{\it Class Lecture:} The instructors need to devote a lecture on some of the basic analog circuits such as threshold comparator, gain amplifier, switching thresholds Schmitt trigger, etc., if such circuits are not yet familiar to the students. Examples of these analog circuits
in the real world would help the students understanding their operation. For example, the work of a threshold comparator can be exemplified by an alarm system in which the alarm is activated when a signal from, e.g., a smoke sensor exceeds a threshold value. It would also be beneficial to discuss the advantages of using memelements in these circuits - such as non-volatile information storage -
rather than other schemes currently employed.

{\it Laboratory Experiment:} Programmable analog circuits proposed in this experiment are based on the threshold-type switching behavior of memory circuit elements \cite{pershin10d}.
Such a behavior is typical for many memristive devices and can be described by the following simple model suggested by the present authors~\cite{pershin09b}:
\begin{eqnarray}
R&=&x, \label{eq43} \\ \dot x&=&\left(\beta V_M+0.5\left(
\alpha-\beta\right)\left[ |V_M+V_t|-|V_M-V_t| \right]\right)
\nonumber\\ & &\times \theta\left( x-R_1\right) \theta\left(
R_2-x\right) \label{eq44},
\end{eqnarray}
where $\theta(\cdot)$ is the step function, $\alpha$  and $\beta$
characterize the rate of memristance change at $|V_M|\leq V_t$ and
$|V_M|> V_t$, respectively, $V_M$ is the voltage drop on memristive system,
$V_t$ is a threshold voltage, and  $R_1$ and $R_2$ are limiting
values of the memristance $R$. In Eq. (\ref{eq44}), the $\theta$-functions
symbolically show that the memristance can change only between
$R_1$ and $R_2$.

Although the previous demonstration of programmable analog circuits was based on memristive systems \cite{pershin10d}, memcapacitive and meminductive systems can be similarly employed in analog circuits. Specifically, this application is based on applying only small signal amplitudes to memory circuit elements in the analog mode of operation - to keep their states unchanged - and use high signal amplitudes to perform the switching. It is convenient to apply the control signals in pulses.

In Ref. \cite{pershin10d}, we have demonstrated several programmable analog circuits with memristive systems including a programmable threshold comparator, programmable gain amplifier, programmable switching thresholds Schmitt trigger (see Fig. \ref{fig3}) and programmable frequency relaxation oscillator. Some of these circuits as well as similar circuits with memcapacitive and meminductive systems can be included in the project. These projects require memristor, memcapacitor and meminductor emulators, pulse generator, power supply and data acquisition system.

\subsection{Associative memory}

{\it Class Lecture:} This is another project where the students can learn that memelements can find applications in the modeling of biological processes. For instance, it is presently accepted that biological synapses play the major role in brain memory and information processing. Analogously, in artificial neural networks \cite{smith10a}, memristive devices can be used as artificial synapses \cite{snider08a,erokhin07a,Choi09a,jo10a,pershin10c,Alibart10a,Lai10a,zhao10a}, namely, as electronic analogs of biological synapses connecting neuron cells.  The instructors should then discuss the known functions of
biological synapses and make a parallel with the properties of memristive systems. The instructors may also want to introduce the students to Hebbian and Pavlonian learning, spike-timing dependent plasticity, and possibly the Hodgkin and Huxley model of channel conductance of nerve membranes \cite{hodgkin52a}.

{\it Laboratory Experiment:} This project demonstrates the electronic simulation of associative memory (see Fig. \ref{fig4} and Ref. \cite{pershin10c} for more details) similar to the famous Pavlov's experiments on conditioned reflexes \cite{pavlov27a} whereby salivation of a dog's mouth is first set by the sight of food. Then, if the sight of food is accompanied by a sound (e.g., the tone of a bell) over a certain period of time, the dog learns to associate the sound to the food, and salivation can be triggered by the sound alone, without the intervention of vision.

The goal of this project would be to construct an artificial neural network to demonstrate the development of associative memory. The simplest Hebbian learning rule should be first demonstrated as in Ref. \cite{pershin10c}. However, a more challenging project would require the modeling of synapses satisfying spike-timing dependent plasticity as in Ref. \cite{pershin10e}. In this case, particular care has to be applied to how incoming and outgoing signals are processed by the memristive systems. Finally, as an additional project the student can be asked to come up with schemes that allow "un-learning", namely allow the associative memory to decay if the sight-of-food signal is not fed to the circuit after a certain
period of time. These projects require two memristor emulators, three electronic neurons (which can be easily constructed using microcontrollers, see Ref. \cite{pershin10c}) and data acquisition system.

\subsection{Logic}

\begin{figure}[t]
 \begin{center}
 \includegraphics[angle=0,width=6cm]{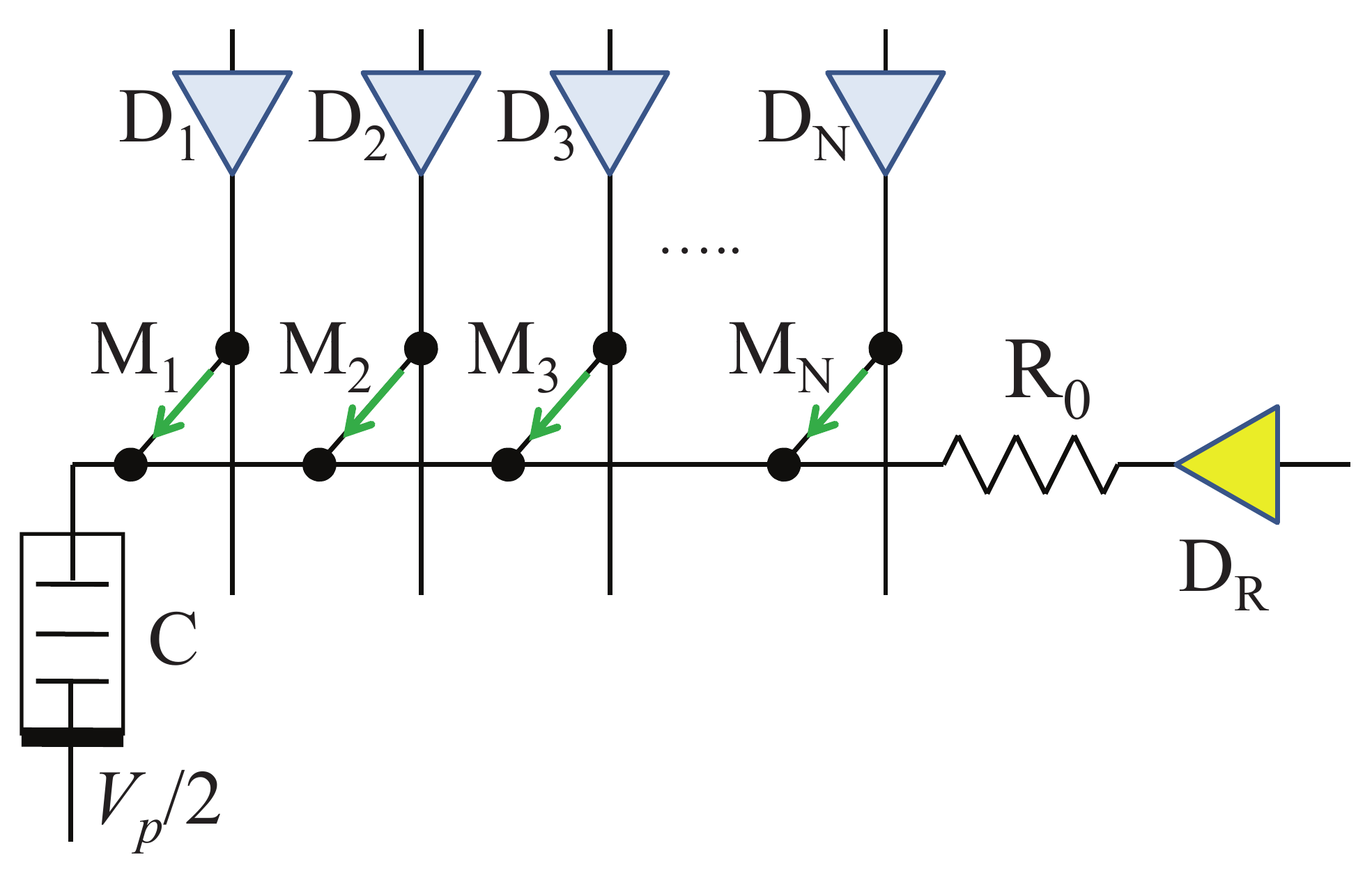}
\caption{\label{fig6} Logic and arithmetic
operations with memory circuit elements. In this circuit, an array of $N$ memristive systems $M_i$, memcapacitive
system $C$ and resistor $R_0$ are connected to a common (horizontal)
line. The circuit operation involves charging the memcapacitive system $C$
through input memristive systems and discharging it through the output ones.
Reprinted with permission from \cite{pershin10e}. Copyright © (2011) by the IEEE.
}
 \end{center}
\end{figure}

{\it Class Lecture:} There is a potentially important application of memory circuit elements in the area of logic \cite{strukov05a,Snider07a,Lehtonen09a,borghetti10a,pershin10e}. "Stateful" logic operations with memristive devices for which the same devices play simultaneously the role of latch and gate were demonstrated experimentally \cite{borghetti10a}. Mathematical operations with memristive systems used as in traditional computers, however, would require a large number of computational steps \cite{Lehtonen09a}. Recently, we have suggested an optimized circuit architecture (see Fig. \ref{fig6}) with memristive systems and
(mem)capacitors that significantly reduces the mathematical operations needed to perform logic \cite{pershin10e}.

The instructors should introduce the students to the basics of boolean logic and arithmetic operations in the
binary system. Also, an important point should be made that in traditional computing systems, information processing and memory storage belong to physically disjoint platforms. The instructors would then introduce the
concept that with memelements one can perform information processing {\it and} memory storage on the {\it same} platform. This lecture would then introduce the students to the impact these elements could have in more
traditional computing paradigms.

{\it Laboratory Experiment:} This project involves the demonstration of basic logic operations - NOT, AND, OR - as well as addition of one-bit numbers with memory circuit elements. The students can either implement the circuit as shown in
Fig.~\ref{fig6} or replace the memcapacitors with regular capacitors. The project requires four memristor emulators, one memcapacitor emulator, voltage source, programmable pulse generator and data acquisition system.

\subsection{Cross-bar memory}

\begin{figure}[tb]
 \begin{center}
 \includegraphics[angle=0,width=6cm]{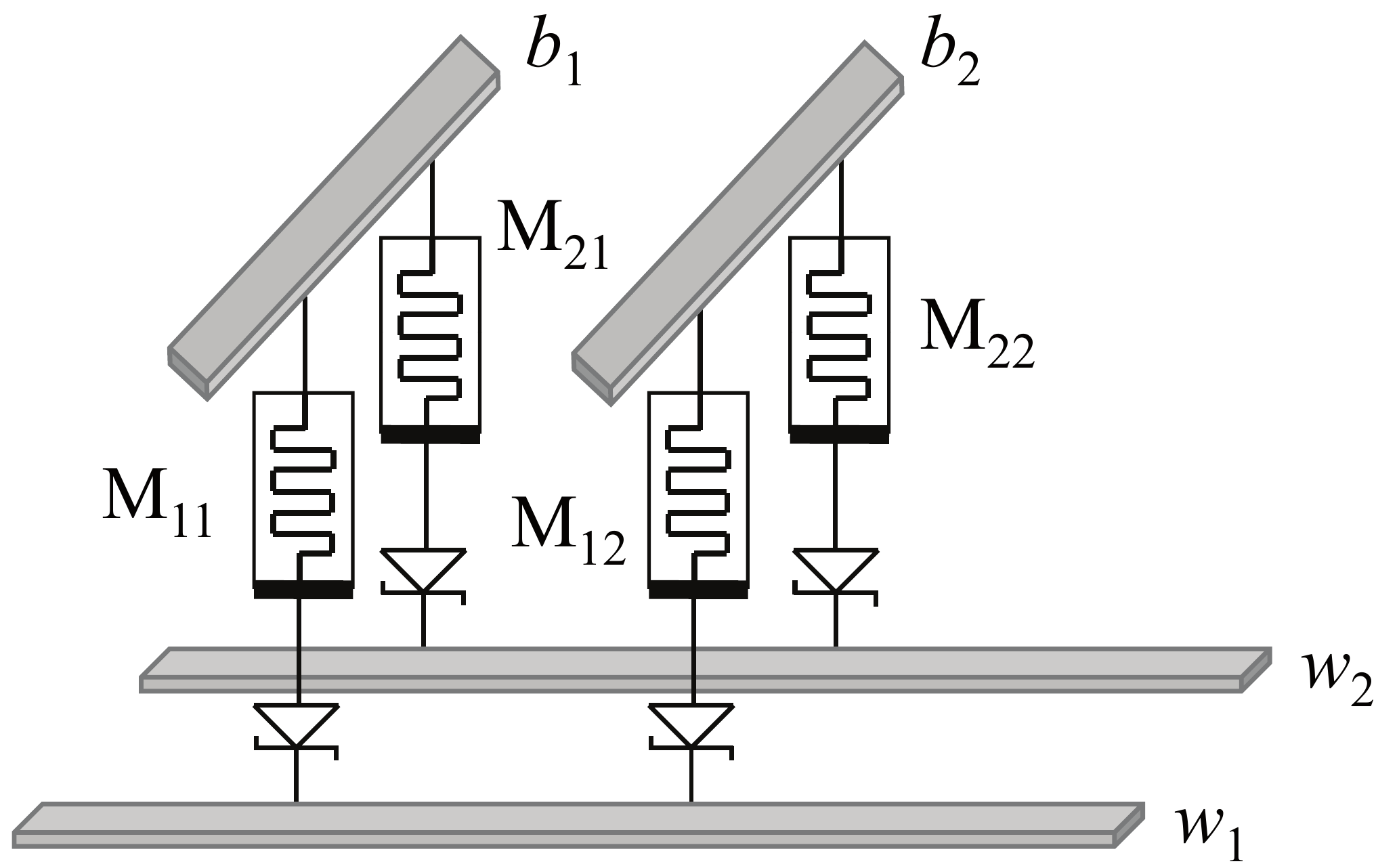}
\caption{\label{fig5} Electronic circuit simulating a cross-bar memory array. Here, word ($w_i$) and bit ($b_i$) lines
are connected by units consisting of a memristive system and a non-linear element (Zener diode). Such an arrangement solves the "sneak
path" problem.}
 \end{center}
\end{figure}

{\it Class Lecture:} The cross-bar array memory structure is very promising for high-density, random access 3D stackable memristive memory \cite{ITRS09a}. It is thus important to introduce such structure to the students as well as to discuss its pros and cons. A single layer of cross-bar array consists of a memristive material sandwiched between two arrays of metal wires perpendicular to each other - so called word and bit lines \cite{pershin11a}. Each bit of information in the cross-bar array is addressed by selection of a pair of orthogonal wires. Thus, an $N\times N$ cross-bar array encodes $N^2$ bits of information. The important problem currently facing the practical implementation of cross-bar memory is the "sneak path" problem~\cite{linn10a}. When a voltage is applied to a particular pair of word and bit wires to perform a read or write operation, the electric current is distributed between both the selected intersection and all other non-selected intersections of word and bit lines. In this way, the whole cross-bar array can be generally seen as a large $N\times N$ resistive network. The instructors should
introduce this problem to the students and present them some solutions (such as the one in Fig. \ref{fig5}). They should also indicate in detail the differences between the 2D and 3D versions of the cross-bar array, including the pros and cons of both.

{\it Laboratory Experiment:} The setup presented in Fig. \ref{fig5} introduces the concept of cross-bar memory arrays as well as a potential solution of the "sneak path" problem via utilization of non-linear elements - in the present case, of Zener diodes. In particular, students can be asked to investigate read and write operations in the cross-bar array with and without Zener diodes. Without diodes, the value of memristance during a reading is distorted by the presence  of "sneak paths". For the same reason, writing operations can alter states of non-selected memristive systems. Students could be asked to demonstrate and explain how the introduction of Zener diodes solves this problem. In addition, they can also be asked to find appropriate circuit operation parameters such as read/write voltage windows to better understand the circuit operation.

A different possible solution of the "sneak path" problem is based on the concept of {\it complementary resistive switches} \cite{linn10a} that can also be studied within this project. For the circuit presented in
Fig.~\ref{fig5} this project requires four memristor emulators, four Zener diodes, power supply and data acquisition system.

\section{Final remarks on the proposed teaching approach} \label{concl}

We are convinced that the most efficient method to teach memory circuit elements consists in an integration of
hands-on experimental labs with a theoretical course. The sequence of laboratory projects we suggest thus provide a useful practical starting point for such an integrated approach.
The projects described in this paper are clearly just representative of a wide range of possible applications of memory circuit elements. Therefore, they are only intended as an inspiration for more case studies. We hope, and indeed expect the instructors to take advantage of these suggestions and significantly extend the range of these projects. As a result, students will learn both the basic principles of operation of memory circuit elements
and get familiar with the wide range of important applications they can be useful for. It is our hope that the rapidly growing field of memelements will then be part of standard university curricula in Physics and Engineering.

\section*{Acknowledgment}
M.D. acknowledges partial support from the National Science
Foundation (DMR-0802830).

\bibliographystyle{IEEEtran}
\bibliography{IEEEabrv,maze}

\begin{thebibliography}{10}
\providecommand{\url}[1]{#1}
\csname url@samestyle\endcsname
\providecommand{\newblock}{\relax}
\providecommand{\bibinfo}[2]{#2}
\providecommand{\BIBentrySTDinterwordspacing}{\spaceskip=0pt\relax}
\providecommand{\BIBentryALTinterwordstretchfactor}{4}
\providecommand{\BIBentryALTinterwordspacing}{\spaceskip=\fontdimen2\font plus
\BIBentryALTinterwordstretchfactor\fontdimen3\font minus
  \fontdimen4\font\relax}
\providecommand{\BIBforeignlanguage}[2]{{%
\expandafter\ifx\csname l@#1\endcsname\relax
\typeout{** WARNING: IEEEtran.bst: No hyphenation pattern has been}%
\typeout{** loaded for the language `#1'. Using the pattern for}%
\typeout{** the default language instead.}%
\else
\language=\csname l@#1\endcsname
\fi
#2}}
\providecommand{\BIBdecl}{\relax}
\BIBdecl

\bibitem{diventra09a}
M.~{Di Ventra}, Y.~V. Pershin, and L.~O. Chua, ``Circuit elements with memory:
  Memristors, memcapacitors, and meminductors,'' \emph{Proc. IEEE}, vol.~97,
  no.~10, pp. 1717--1724, 2009.

\bibitem{chua71a}
L.~O. Chua, ``Memristor - the missing circuit element,'' \emph{IEEE Trans.
  Circuit Theory}, vol.~18, pp. 507--519, 1971.

\bibitem{chua76a}
L.~O. Chua and S.~M. Kang, ``Memristive devices and systems,'' \emph{Proc.
  IEEE}, vol.~64, pp. 209--223, 1976.

\bibitem{pershin11a}
Y.~V. Pershin and M.~Di~Ventra, ``Memory effects in complex materials and
  nanoscale systems,'' \emph{Advances in Physics}, vol.~60, pp. 145--227, 2011.

\bibitem{Green07a}
J.~E. Green, J.~W. Choi, A.~Boukai, Y.~Bunimovich, E.~Johnston-Halperin,
  E.~DeIonno, Y.~Luo, B.~A. Sheriff, K.~Xu, Y.~S. Shin, H.-R. Tseng, J.~F.
  Stoddart, and J.~R. Heath, ``A 160-kilobit molecular electronic memory
  patterned at 10$^{11}$ bits per square centimetre,'' \emph{Nature}, vol. 445,
  p. 414, 2007.

\bibitem{Karg08a}
S.~F. Karg, G.~I. Meijer, J.~G. Bednorz, C.~T. Rettner, A.~G. Schrott, E.~A.
  Joseph, C.~H. Lam, M.~Janousch, U.~Staub, F.~La~Mattina, S.~F. Alvarado,
  D.~Widmer, R.~Stutz, U.~Drechsler, and D.~Caimi,
  ``Transition-metal-oxide-based resistance-change memories,'' \emph{IBM J.
  Res. Dev.}, vol.~52, no. 4-5, pp. 481--492, JUL-SEP 2008.

\bibitem{Sawa08a}
A.~Sawa, ``Resistive switching in transition metal oxides,'' \emph{Mat. Today},
  vol.~11, pp. 28--36, 2008.

\bibitem{pershin11d}
Y.~V. Pershin and M.~Di~Ventra, ``Solving mazes with memristors: a
  massively-parallel approach,'' \emph{Phys. Rev. E}, vol.~84, p. 046703, 2011.

\bibitem{pershin10c}
------, ``Experimental demonstration of associative memory with memristive
  neural networks,'' \emph{Neural {N}etworks}, vol.~23, p. 881, 2010.

\bibitem{jo10a}
S.~H. Jo, T.~Chang, I.~Ebong, B.~B. Bhadviya, P.~Mazumder, and W.~Lu,
  ``Nanoscale memristor device as synapse in neuromorphic systems,'' \emph{Nano
  Lett.}, vol.~10, pp. 1297--1301, 2010.

\bibitem{Choi09a}
H.~Choi, H.~Jung, J.~Lee, J.~Yoon, J.~Park, D.-J. Seong, W.~Lee, M.~Hasan,
  G.-Y. Jung, and H.~Hwang, ``An electrically modifiable synapse array of
  resistive switching memory,'' \emph{Nanotechn.}, vol.~20, p. 345201, 2009.

\bibitem{Lai10a}
Q.~Lai, L.~Zhang, Z.~Li, W.~F. Stickle, R.~S. Williams, and Y.~Chen,
  ``Ionic/electronic hybrid materials integrated in a synaptic transistor with
  signal processing and learning functions,'' \emph{Adv. Mat.}, vol.~22, p.
  2448, 2010.

\bibitem{Alibart10a}
F.~Alibart, S.~Pleutin, D.~Guerin, C.~Novembre, S.~Lenfant, K.~Lmimouni,
  C.~Gamrat, and D.~Vuillaume, ``An organic nanoparticle transistor behaving as
  a biological spiking synapse,'' \emph{Adv. Funct. Mat.}, vol.~20, pp.
  330--337, 2010.

\bibitem{fontana10a}
M.~P. Fontana, 2010, private communication.

\bibitem{strukov05a}
D.~Strukov and K.~Likharev, ``{CMOL FPGA: a reconfigurable architecture for
  hybrid digital circuits with two-terminal nanodevices},''
  \emph{{Nanotechn.}}, vol.~{16}, pp. {888--900}, {2005}.

\bibitem{Snider07a}
G.~S. Snider and R.~S. Williams, ``Nano/{CMOS} architectures using a
  field-programmable nanowire interconnect,'' \emph{Nanotechnology}, vol.~18,
  p. 035204, 2007.

\bibitem{borghetti10a}
J.~Borghetti, G.~S. Snider, P.~J. Kuekes, J.~J. Yang, D.~R. Stewart, and R.~S.
  Williams, ``{`Memristive' switches enable `stateful' logic operations via
  material implication},'' \emph{Nature}, vol. {464}, pp. {873--876}, {2010}.

\bibitem{pershin10e}
\BIBentryALTinterwordspacing
Y.~V. Pershin and M.~{Di Ventra}, ``Neuromorphic, digital and quantum
  computation with memory circuit elements,'' \emph{Proc. {IEEE} (in press);
  ar{X}ive:1009.6025}, 2012. [Online]. Available:
  \url{http://dx.doi.org/10.1109/JPROC.2011.2166369}
\BIBentrySTDinterwordspacing

\bibitem{pershin09b}
Y.~V. Pershin, S.~{La Fontaine}, and M.~{Di Ventra}, ``Memristive model of
  amoeba learning,'' \emph{Phys. Rev. E}, vol.~80, p. 021926, 2009.

\bibitem{Johnsen11a}
G.~K. Johnsen, C.~A. L\"utken, O.~G. Martinsen, and S.~Grimnes, ``Memristive
  model of electro-osmosis in skin,'' \emph{Phys. Rev. E}, vol.~83, p. 031916,
  2011.

\bibitem{reram2013}
Announced by Stan Williams, HP, at IEF2011 meeting in Seville, Spain, October
  2011.

\bibitem{pershin10d}
Y.~V. Pershin and M.~Di~Ventra, ``Practical approach to programmable analog
  circuits with memristors,'' \emph{IEEE Trans. Circ. Syst. I}, vol.~57, p.
  1857, 2010.

\bibitem{pershin11b}
------, ``Emulation of floating memcapacitors and meminductors using current
  conveyors,'' \emph{Electronics Letters}, vol.~47, p. 243, 2011.

\bibitem{Pazienza11a}
J.~Albo-Canals and G.~E. Pazienza, ``How to teach memristors in ee
  undergraduate courses,'' in \emph{Proceedings of the 2011 International
  Symposium on {Circuits and Systems (ISCAS)}}, 2011, pp. 345 -- 348.

\bibitem{pershin09e}
Y.~V. Pershin and M.~Di~Ventra, ``Memristive circuits simulate memcapacitors
  and meminductors,'' \emph{Electronics Letters}, vol.~46, pp. 517--518, 2010.

\bibitem{biolek10a}
D.~Biolek and V.~Biolkova, ``Mutator for transforming memristor into
  memcapacitor,'' \emph{Ell. Lett.}, vol.~46, p. 1428, 2010.

\bibitem{biolek11a}
D.~Biolek, V.~Biolkova, and Z.~Kolka, ``Low-voltage-low- power current conveyor
  for battery supplied memristor emulator,'' in \emph{In 5th International
  Conference on Circuits, Systems and Signals ({CSS'11}).}, 2011, p. 171.

\bibitem{biolek11b}
D.~Biolek and V.~Biolkova, ``Hybrid modelling and emulation of mem-systems,''
  \emph{Int. J. Numer. Mod. (in press)}, 2011.

\bibitem{Wang2011a}
\BIBentryALTinterwordspacing
X.~Wang, A.~Fitch, H.~Iu, and W.~Qi, ``Design of a memcapacitor emulator based
  on a memristor,'' \emph{Physics Letters A (in press)}, 2011. [Online].
  Available:
  \url{http://www.sciencedirect.com/science/article/pii/S037596011101365X}
\BIBentrySTDinterwordspacing

\bibitem{webpage}
Technical documentation on memristor emulator can be found at
  \url{http://www.physics.sc.edu/%7epershin/emulator.htm}.

\bibitem{yang08a}
J.~J. Yang, M.~D. Pickett, X.~Li, D.~A.~A. Ohlberg, D.~R. Stewart, and R.~S.
  Williams, ``Memristive switching mechanism for metal/oxide/metal
  nanodevices,'' \emph{Nat. Nanotechnol.}, vol.~3, pp. 429--433, 2008.

\bibitem{jo09a}
S.~H. Jo, K.-H. Kim, and W.~Lu, ``High-density crossbar arrays based on a si
  memristive system,'' \emph{Nano Lett.}, vol.~9, pp. 870--874, 2009.

\bibitem{waser07a}
R.~R. Waser and M.~Aono, ``Nanoionics-based resistive switching memories,''
  \emph{Nat. mat.}, vol.~6, pp. 833--840, 2007.

\bibitem{martinez09a}
J.~Martinez-Rincon, M.~Di~Ventra, and Y.~V. Pershin, ``Solid-state
  memcapacitive system with negative and diverging capacitance,'' \emph{Phys.
  Rev. B}, vol.~81, p. 195430, 2010.

\bibitem{pershin08a}
Y.~V. Pershin and M.~{Di Ventra}, ``Spin memristive systems: Spin memory
  effects in semiconductor spintronics,'' \emph{Phys. Rev. B}, vol.~78, pp.
  113\,309--1--113\,309--4, 2008.

\bibitem{wang09a}
X.~Wang, Y.~Chen, H.~Xi, H.~Li, and D.~Dimitrov, ``Spintronic memristor through
  spin-torque-induced magnetization motion,'' \emph{El. Dev. Lett.}, vol.~30,
  pp. 294 -- 297, 2009.

\bibitem{pershin11c}
J.~Martinez-Rincon and Y.~V. Pershin, ``Bistable non-volatile elastic membrane
  memcapacitor exhibiting chaotic behavior,'' \emph{IEEE Trans. El. Dev.},
  vol.~58, p. 1809, 2011.

\bibitem{Wright11a}
C.~D. Wright, Y.~Liu, K.~I. Kohary, M.~M. Aziz, and R.~J. Hicken, ``Arithmetic
  and biologically-inspired computing using phase-change materials,''
  \emph{Advanced Materials}, vol.~23, pp. 3408--3413, 2011.

\bibitem{krems2010a}
M.~Krems, Y.~V. Pershin, and M.~Di~Ventra, ``Ionic memcapacitive effects in
  nanopores,'' \emph{Nano Lett.}, vol.~10, p. 2674, 2010.

\bibitem{Corinto11a}
F.~Corinto, A.~Ascoli, and M.~Gilli, ``Analysis of current-voltage
  characteristics for memristive elements in pattern recognition systems,''
  \emph{submitted for publication}, 2011.

\bibitem{Stotland11a}
A.~Stotland and M.~Di~Ventra, ``Stochastic memory: getting memory out of
  noise,'' \emph{arXiv:1104.4485v2}, 2011.

\bibitem{driscoll11a}
T.~Driscoll, Y.~V. Pershin, D.~N. Basov, and M.~{Di Ventra}, ``Chaotic
  memristor,'' \emph{Applied Physics A}, vol. 102, p. 885, 2011.

\bibitem{saigusa08a}
T.~Saigusa, A.~Tero, T.~Nakagaki, and Y.~Kuramoto, ``Amoebae anticipate
  periodic events,'' \emph{Phys. Rev. Lett.}, vol. 100, p. 018101, 2008.

\bibitem{smith10a}
L.~S. Smith, ``Neuromorphic systems: Past, present and future,'' in \emph{Brain
  inspired cognitive systems 2008}, ser. Advances in Experimental Medicine and
  Biology, A.~Hussain, I.~Aleksander, L.~Smith, A.~Barros, R.~Chrisley, and
  V.~Cutsuridis, Eds., 2010, vol. 657, pp. 167--182.

\bibitem{snider08a}
G.~S. Snider, ``Cortical computing with memristive nanodevices,'' \emph{SciDAC
  Review}, vol.~10, pp. 58--65, 2008.

\bibitem{erokhin07a}
V.~V. Erokhin, T.~S. Berzina, and M.~P. Fontana, ``Polymeric elements for
  adaptive networks,'' \emph{Crystallogr. Rep.}, vol.~52, pp. 159--166, 2007.

\bibitem{zhao10a}
W.~S. Zhao, G.~Agnus, V.~Derycke, A.~Filoramo, J.-P. Bourgoin, and C.~Gamrat,
  ``Nanotube devices based crossbar architecture: toward neuromorphic
  computing,'' \emph{Nanotechnology}, vol.~21, no.~17, p. 175202, 2010.

\bibitem{hodgkin52a}
A.~L. Hodgkin and A.~F. Huxley, ``A quantitative description of membrane
  current and its application to conduction and excitation in nerve,''
  \emph{Journal of Physiology}, vol. 117, pp. 500--544, 1952.

\bibitem{pavlov27a}
I.~Pavlov, \emph{Conditioned Reflexes: An Investigation of the Physiological
  Activity of the Cerebral Cortex}.\hskip 1em plus 0.5em minus 0.4em\relax
  London: Oxford University Press, 1927, (translated by G. V. Anrep).

\bibitem{Lehtonen09a}
E.~Lehtonen and M.~Laiho, ``Stateful implication logic with memristors,'' in
  \emph{Proceedings of the 2009 International Symposium on {Nanoscale
  Architectures (NANOARCH'09))}}, 2009, pp. 33 -- 36.

\bibitem{ITRS09a}
2009, {ITRS}. The International Technology Roadmap for Semiconductors - ITRS
  2009 Edition. http://www.itrs.net.

\bibitem{linn10a}
E.~Linn, R.~Rosezin, and R.~Waser, ``Complementary resistive switches for
  passive nanocrossbar memories,'' \emph{Nature Mat.}, vol.~9, p. 403, 2010.

\end{thebibliography}
\end{document}